\documentclass[pdflatex,sn-mathphys-num]{sn-jnl}

\usepackage{graphicx}%
\usepackage{multirow}%
\usepackage{amsmath,amssymb,amsfonts}%
\usepackage{amsthm}%
\usepackage{mathrsfs}%
\usepackage[title]{appendix}%
\usepackage{xcolor}%
\usepackage[normalem]{ulem}%
\usepackage{textcomp}%
\usepackage{manyfoot}%
\usepackage{booktabs}%
\usepackage{algorithm}%
\usepackage{algorithmicx}%
\usepackage{algpseudocode}%
\usepackage{listings}%
\usepackage{diagbox}

\theoremstyle{thmstyleone}%
\theoremstyle{thmstyletwo}%

\theoremstyle{thmstylethree}%

\begin{document}

\title[Article Title]{EPIC-CIM: Training Convolutional Neural Networks on a Coherent Ising Machine via Equilibrium Propagation}

\author[1]{\fnm{Xingrui} \sur{Yin}}

\equalcont{These authors contributed equally to this work.}

\author[2]{\fnm{Shenwei} \sur{Kang}}

\equalcont{These authors contributed equally to this work.}

\author[1]{\fnm{Haoqi} \sur{He}}

\author[1]{\fnm{Yan} \sur{Xiao}}
\email{xiaoy367@mail.sysu.edu.cn}

\author[2]{\fnm{Hongdong} \sur{Zhu}}

\author[2]{\fnm{Hai} \sur{Wei}}

\author[2]{\fnm{Yin} \sur{Ma}}

\author[2]{\fnm{Qi} \sur{Gao}}
\email{gaoq@boseq.com}

\author[1]{\fnm{Xiaochun} \sur{Cao}}

\author[2]{\fnm{Kai} \sur{Wen}}
\email{wenk@boseq.com}

\affil*[1]{%
\orgdiv{},
\orgname{Sun Yat-sen University},
\orgaddress{%
\city{Guangzhou},
\country{China}
}}

\affil[2]{%
\orgdiv{},
\orgname{QBoson Quantum Technology Co., Ltd.},
\orgaddress{%
\city{Beijing},
\country{China}
}}


\abstract{
Quantum convolutional neural networks, due to the involvement of quantum measurements and discrete quantum state evolution, face inherent training challenges associated with non-differentiable operations and discrete optimization dynamics, which make conventional gradient-based learning difficult to apply effectively. In this context, energy-based learning provides a promising alternative by reformulating network training as an energy minimization process without explicit gradient backpropagation.
In this framework, input data are processed through convolutional operations, followed by quantum sampling to generate intermediate binary representations, while the output layer also relies on quantum sampling to produce final predictions. The overall network energy is composed of convolutional feature matching terms, linear coupling terms at the output layer, and global output constraint terms, enabling both parameter updates and feature evolution to be described through physically interpretable energy dynamics. Furthermore, under the equilibrium propagation mechanism, the energy difference between the free phase and the weakly clamped phase is exploited to drive parameter updates without explicit gradient computation, thereby enabling stable and consistent learning in non-differentiable and discrete spaces. While remaining consistent with classical convolutional learning theory, the proposed framework enhances interpretability and observability through quantum energy modeling, offering a unified physical perspective for efficient QCNN training and the integration of quantum computing with artificial intelligence.}


\maketitle

\section{Introduction}\label{sec1}

Deep neural networks~\cite{bib1} have achieved remarkable success in computer vision, natural language processing~\cite{bib2}, and other domains, largely due to gradient-based optimization through backpropagation~\cite{bib3}. However, as network architectures increasingly evolve toward discrete, binarized, or physically implementable systems~\cite{bib4,bib5}, traditional gradient methods~\cite{bib15,bib24} face inherent challenges in handling non-differentiable operators, discrete variables~\cite{bib32}, and complex energy landscapes. These limitations not only reduce training stability but also obscure the physical interpretability of parameter updates, motivating a reconsideration of neural network optimization from the perspective of statistical physics and energy minimization~\cite{bib6,bib7}.

Ising models and their quadratic unconstrained binary optimization (QUBO) formulations provide a unified mathematical framework for modeling discrete energy evolution, and have demonstrated advantages in combinatorial optimization and quantum computing~\cite{bib8,bib9,bib10}. Variational quantum circuits (VQC)~\cite{bib33} have been widely employed to train quantum neural networks on benchmark datasets such as MNIST~\cite{bib11}. While these methods achieve limited success in small-scale or partial classification tasks, they are constrained by circuit depth, noise~\cite{bib34}, and gradient instability~\cite{bib35}, making it challenging to scale to full-classification settings or large datasets~\cite{bib12,bib13,bib14}. Recent studies have explored simulated~\cite{bib36} or quantum annealing~\cite{bib23} approaches, mapping neural network training to Ising or QUBO energy minimization. These approaches have successfully supported full-classification tasks for fully connected networks~\cite{bib15,bib16}. However, extending them to convolutional architectures remains underexplored, as hardware connectivity, binary variable scaling~\cite{bib29}, and efficient representation of local spatial features impose strict limitations on the size of trainable images.

The convolutional neural network (CNN) architecture has been highly successful in classical vision tasks due to its local receptive field and weight-sharing mechanisms~\cite{bib17}. Incorporating convolutional structures into quantum models, however, introduces multiple challenges. Increased input dimensions rapidly expand the number of required quantum variables~\cite{bib18}, and convolution and pooling operations lack efficient representations within Ising or energy-based frameworks. Consequently, existing quantum convolutional neural networks are mostly theoretical or limited to very small-scale experiments, and have rarely been tested on complete multi-class datasets~\cite{bib19}. These limitations hinder the practical applicability of QCNN in real-world visual tasks~\cite{bib15}.

Coherent Ising machines~\cite{bib20,bib21} provide a promising hardware platform for energy-driven neural network training. Compared with annealing-based quantum devices~\cite{bib23}, they offer higher connectivity and parallel evolution capabilities, enabling the natural implementation of complex energy functions derived from neural network optimization~\cite{bib22}. Our contribution is not simply to replace one Ising solver with another. EPIC-CIM develops a convolutional energy formulation in which local feature extraction, binary latent-state sampling, and output constraints are represented within a single CIM-compatible learning dynamics. This distinguishes the proposed framework from prior Ising-machine~\cite{bib25,bib26} training studies in which convolutional models were either compact topology-matching demonstrations or extensions of more general energy-based networks.

Overall, the proposed framework reconstructs gradient estimation under non-differentiable operators as a physically interpretable energy evolution process, thereby significantly improving training stability and robustness while enhancing the interpretability of parameter updates. It enables a coherent integration of non-gradient supervised learning with a physically grounded energy evolution paradigm, providing a feasible solution for scalable QCNN training in full-class classification tasks. Furthermore, it offers a unified and observable physical perspective for bridging quantum computing and artificial intelligence.

To validate the effectiveness of the proposed framework, extensive experiments are conducted on three representative image classification benchmarks, namely MNIST, Fashion-MNIST, and CIFAR-10. The proposed EPIC-CIM is compared with representative quantum neural network models, including variational quantum circuits , quantum convolutional neural networks , quantum deep equilibrium models , and quantum annealing-based approaches. Experimental results consistently demonstrate that EPIC-CIM achieves superior classification performance across all datasets while exhibiting improved training stability and robustness. In particular, the proposed method maintains high accuracy in full multi-class classification tasks, highlighting its effectiveness in overcoming the optimization difficulties introduced by discrete quantum states and non-differentiable learning dynamics.

The main contributions of this work are summarized as follows. First, we propose EPIC-CIM, an energy-based quantum convolutional neural network training framework that reformulates convolutional learning as a physically interpretable energy minimization process compatible with coherent Ising machines. Second, we design a unified energy formulation that jointly models convolutional feature extraction, binary latent-state evolution, and supervised output constraints, enabling equilibrium propagation to optimize QCNNs without conventional gradient backpropagation. Third, extensive experimental evaluations on multiple benchmark datasets demonstrate that the proposed framework consistently outperforms existing quantum neural network models while providing stable optimization behavior and improved scalability for full multi-class image classification.


\section{Methods}\label{sec2}

The proposed framework consists of four coupled stages. First, a classical convolutional operator extracts local spatial responses from the input image. Second, these responses are mapped to external fields that drive a binary latent representation sampled by a CIM-compatible Ising energy model. Third, the binary latent states are coupled to output variables subject to a one-hot classification constraint. Finally, learning is performed by comparing the equilibrium states obtained in the free and weakly nudged phases, producing parameter updates for the convolutional and output layers through measurable state correlations.

\subsection{QUBO and Ising Energy Formulation}
\label{subsec:qubo_ising}

The Quadratic Unconstrained Binary Optimization model~\cite{bib27,bib28} provides a unified mathematical formulation for mapping binary neural networks into an energy minimization framework. The core idea is to describe the optimization variables using binary variables taking values in \{0,1\}, and to represent the system energy through a symmetric matrix that incorporates all linear terms and quadratic coupling terms~\cite{bib21,bib30}. In the absence of additional constraints, the optimal solution can be obtained by minimizing the energy function.

The energy function of the QUBO model is given by
\begin{equation}
E(\mathbf{x}) = \sum_{i} Q_{ii} x_i + \sum_{i<j} Q_{ij} x_i x_j
= \mathbf{x}^{\mathsf{T}} Q \mathbf{x}
\label{eq:qubo_energy}
\end{equation}

To map the QUBO model to the Ising model~\cite{bib31}, the binary variables $x_i \in \{0,1\}$ are replaced by Ising spin variables $s_i \in \{-1,+1\}$ through the following linear transformation:
\begin{equation}
x_i = \frac{1 + s_i}{2}
\end{equation}
Substituting this relation into the QUBO energy function yields the corresponding Ising formulation
\begin{equation}
E_{\mathrm{Ising}}(\mathbf{s}) = \sum_i h_i s_i + \sum_{i<j} J_{ij} s_i s_j + \mathrm{const}
\label{eq:ising_energy}
\end{equation}
where the coefficients satisfy 
\begin{equation}
J_{ij} = \frac{1}{4} Q_{ij}, \qquad
h_i = \frac{1}{2} Q_{ii} + \frac{1}{4} \sum_{j \neq i} Q_{ij}
\label{eq:ising_coeff}
\end{equation}

This derivation shows that the linear and quadratic terms in the QUBO formulation can be directly mapped to the local fields and spin--spin couplings in the Ising model, respectively, thereby establishing a complete equivalence between the two models in terms of energy minimization. Consequently, any neural network composed of binary variables can be expressed in a standard QUBO or Ising energy representation, allowing network training to be interpreted as a global energy minimization problem and efficiently solved using dedicated optimization hardware such as coherent Ising machines~\cite{bib20}.

\subsection{Energy-Based QCNN Model}
\label{subsec:energy_qcnn}

\begin{figure}[h]
\centering
\includegraphics[width=1\textwidth]{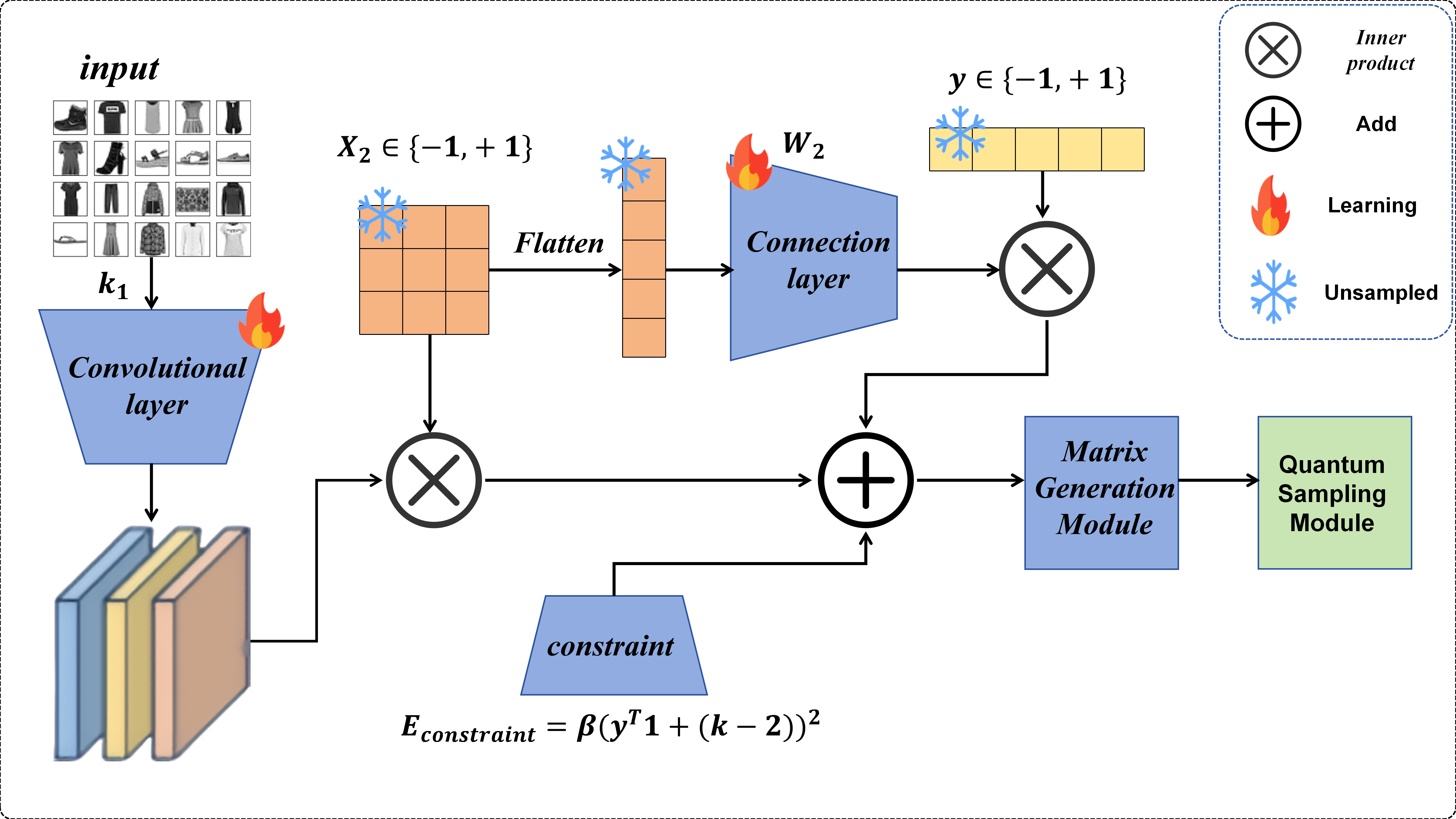}
\caption{Training framework for quantum convolutional neural networks. The model integrates convolutional operations with a quantum sampling module to generate binary intermediate representations. The network is formulated under a unified QUBO/Ising energy model, where both intermediate states and output variables jointly determine the overall energy landscape.}
\label{fig:qcnn_framework}
\end{figure}

Traditional convolutional neural networks  typically rely on gradient-based backpropagation for parameter optimization. However, when discrete variables or non-continuous activation functions are involved, such methods often suffer from non-differentiability and optimization difficulties. To address this issue, we adopt an energy-based modeling perspective, unifying representation learning and parameter optimization within an energy minimization framework. By introducing a quantum sampling mechanism, intermediate feature representations are discretized into binary states.

Specifically, by constraining intermediate features to binary values and interpreting them as the states of a quantum spin system, the forward propagation process can be reformulated as searching for optimal configurations under a given energy function. This formulation naturally captures interactions among discrete features and enables efficient exploration in complex non-convex energy landscapes.

An overview of the proposed QCNN training framework is illustrated in Fig.~\ref{fig:qcnn_framework}. Based on the unified formulation of the QUBO and Ising models, we extend the classical CNN to a quantum convolutional neural network , in which intermediate binary representations are obtained via quantum sampling~\cite{bib37,bib38}. In this model, the input feature map $X_1$ consists of real-valued data. After convolution, it is fed into a quantum sampling module to generate a binary intermediate layer $X_2 \in \{-1,+1\}$, which can be interpreted as the state of a quantum spin system. The output layer $y$ also takes values in $\{-1,+1\}$ and, together with $X_2$, participates in the construction of the network energy function.

 Here, $\mathrm{Conv}(X_1,k_1)$ denotes the vectorized convolutional response obtained by applying the learnable kernel $k_1$ to the input feature map $X_1$, and the inner product with $X_2$ is taken over the corresponding spatial and channel dimensions. The variables $X_2$ and $y$ are binary states sampled from the energy model, whereas $k_1$, $W_2$, and $b$ are trainable parameters. This notation makes explicit how convolutional responses act as external fields for the latent binary states and how the output layer is represented through linear couplings and bias terms. The combination of a convolutional feature-matching term and a linear output coupling term,
\begin{equation}
\begin{aligned}
E
&= \mathrm{Conv}(X_1, k_1) \cdot X_2
+ \bigl(W_2 \cdot \mathrm{flatten}(X_2) + b\bigr) \cdot y 
\end{aligned}
\label{eq:qcnn_energy_base}
\end{equation}
The first term is formed by the convolution of the kernel $k_1$ with the input feature map $X_1$ and represents the driving force or energy contribution from the input layer to the binary quantum intermediate layer $X_2$. The second term describes the linear coupling between the quantum-generated intermediate layer $X_2$ and the output layer $y$.

With this energy definition, both the forward propagation and training processes of the QCNN can be interpreted as minimizing a quantum energy function. The generation of the intermediate layer $X_2$ and the output layer $y$ is directly accomplished via quantum sampling, enabling a tight integration of quantum optimization with the convolutional network structure.

To ensure that the output neurons satisfy a $k$-class one-hot constraint~\cite{bib39,bib40}, namely that only one neuron is activated at the optimal solution while the remaining neurons take negative values, a global constraint term $(y^\top \mathbf{1} + (k - 2))^2$ is introduced. This term induces quadratic couplings in the energy function,
\begin{equation}
E_{\mathrm{constraint}}
= \beta \left( y^\top \mathbf{1} + (k - 2) \right)^2 
\label{eq:qcnn_constraint}
\end{equation}

By combining the base energy function with the global constraint term, the complete energy function of the QCNN is obtained as
\begin{equation}
\begin{aligned}
E_{\mathrm{total}}
&= \mathrm{Conv}(X_1, k_1) \cdot X_2
+ \bigl(W_2 \cdot \mathrm{flatten}(X_2) + b\bigr) \cdot y \\
&\quad + \beta \left( y^\top \mathbf{1} + (k - 2) \right)^2 
\end{aligned}
\label{eq:qcnn_energy_total}
\end{equation}

Under this energy formulation, the forward propagation and training of the QCNN are unified as a quantum energy minimization process: the intermediate layer $X_2$ and the output layer $y$ are generated directly through quantum sampling, the convolutional outputs and residual mapping weights primarily contribute linear terms, while the global output constraint gives rise to quadratic terms. This framework enables a tight integration of quantum optimization mechanisms with convolutional network architectures~\cite{bib41,bib42}.

\subsection{Energy-Based Learning and Equilibrium Propagation}
\label{subsec:energy_ep}

Under the above energy-based modeling framework, both the forward propagation and parameter learning of neural networks can be uniformly formulated as optimization problems over an energy function. In this work, we consider a supervised learning task based on the mean squared error (MSE) loss~\cite{bib49}, where the network parameters are discretely quantized variables and are optimized using an Ising machine.

For a given dataset, the training process of a multilayer QCNN can be formalized as the following optimization problem:
\begin{equation}
\theta^\ast
= \arg\min_{\theta}
\frac{1}{N}
\sum_{i=1}^{N}
\left( y_i - f(x_i; \theta) \right)^2 
\label{eq:erm_mse}
\end{equation}
where $f(x_i; \theta)$ denotes a multilayer feedforward network with quantized parameters $\theta$.

\subsection{Energy-Based QCNN Model}
\label{subsec:energy_qcnn}

\begin{figure}[h]
\centering
\includegraphics[width=1\textwidth]{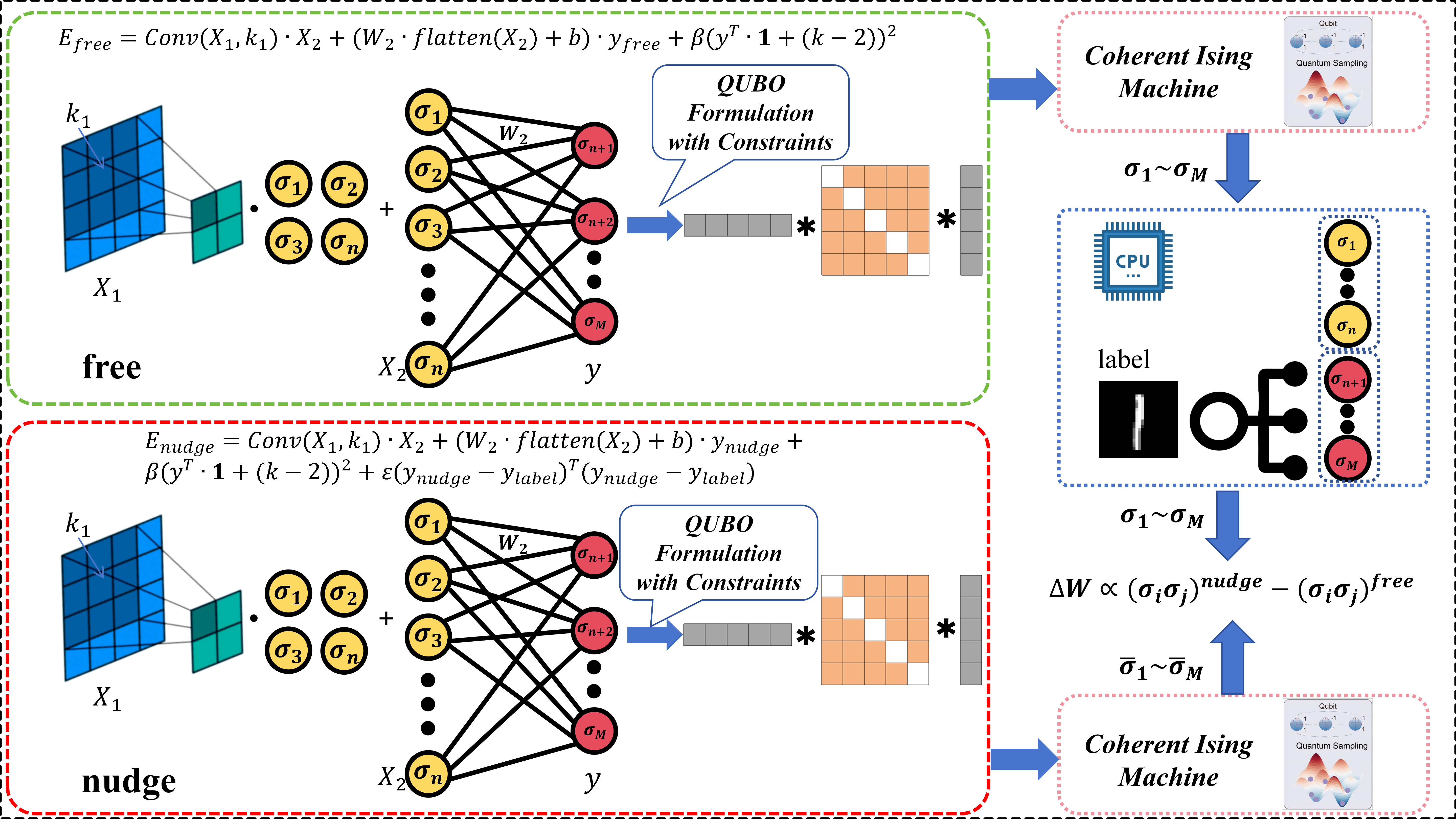}
\caption{Equilibrium Propagation Algorithm Diagram.The figure illustrates the learning process of equilibrium propagation by comparing the system dynamics in the free phase and the weakly nudged phase. Parameter updates are driven by the difference between the two equilibrium states under the same energy function. This mechanism enables gradient estimation through local state variations without explicit backpropagation.}
\label{fig:ep_mechanism}
\end{figure}

Since the parameter space is discrete, the above empirical risk minimization problem can be further mapped to an equivalent Ising or QUBO energy minimization formulation, which can then be efficiently solved using quantum or quantum-inspired Ising machines. It is worth noting that this energy-minimization-based Ising learning framework exhibits strong generality: beyond supervised learning, it can be naturally extended to a variety of learning paradigms, including self-supervised learning and reinforcement learning~\cite{bib50,bib51}.

Equilibrium Propagation~\cite{bib24} is a supervised learning algorithm for physical systems proposed by Scellier and Bengio in 2017~\cite{bib43,bib44}. It aims to enable efficient training of energy-based models through local learning rules, with parameter updates that are consistent with gradient-based methods such as backpropagation at the first-order approximation level. EP unifies inference and learning within the energy evolution of the system.

In the free phase, the input is clamped while the output remains unconstrained, allowing the system to naturally evolve toward a steady state. The corresponding energy function can be written as

\begin{equation}
\begin{aligned}
E_{\mathrm{free}}
&= E(\theta, x, y_{\mathrm{free}}) \\
&= \mathrm{Conv}(x, k_1) \cdot X_2
+ \bigl(W_2 \cdot \mathrm{flatten}(X_2) + b\bigr) \cdot y_{\mathrm{free}} \\
&\quad + \beta \left( y^\top \mathbf{1} + (k - 2) \right)^2
\end{aligned}
\label{eq:free_energy}
\end{equation}

where $x$ denotes the input, $\theta$ represents the system parameters, and $y_{\mathrm{free}}$ is the steady-state output obtained in the free phase.

In the nudged phase, a small perturbation $\epsilon$ is applied to the output units to bias the system state toward the target output $y_{\mathrm{label}}$. The energy function is then given by

\begin{equation}
\begin{aligned}
E_{\mathrm{nudge}}
&= E(\theta, x, y_{\mathrm{nudge}})
+ \epsilon\, C(y_{\mathrm{nudge}}, y_{\mathrm{label}}) \\
&= \mathrm{Conv}(x, k_1) \cdot X_2
+ \bigl(W_2 \cdot \mathrm{flatten}(X_2) + b\bigr) \cdot y_{\mathrm{nudge}} \\
&\quad + \beta \left( y^\top \mathbf{1} + (k - 2) \right)^2
+ \epsilon (y_{\mathrm{nudge}} - y_{\mathrm{label}})^\top
(y_{\mathrm{nudge}} - y_{\mathrm{label}})
\end{aligned}
\label{eq:nudge_energy}
\end{equation}

where $C(\cdot)$ denotes the loss function and $\epsilon$ controls the strength of the output perturbation.

In conventional quantum circuit neural networks, model parameters are typically updated using gradient-based backpropagation. The parameter update rule can be generally expressed as
\begin{equation}
\Delta \theta \propto 
\frac{\partial E_{\mathrm{free}}}{\partial \theta} - 
\frac{\partial E_{\mathrm{nudge}}}{\partial \theta}.
\end{equation}

Under the equilibrium propagation  framework, by comparing the equilibrium states reached in the free phase and the weakly clamped (nudged) phase, the parameter update can be reformulated using only locally measurable quantities, without explicit backpropagation. Specifically, the update rule can be written as
\begin{equation}
\Delta \theta = \frac{1}{\beta} \left[
(\sigma_i \sigma_j)^{\mathrm{nudge}} - 
(\sigma_i \sigma_j)^{\mathrm{free}}
\right].
\end{equation}


From the perspective of energy-based learning, EP is conceptually related to Contrastive Divergence (CD)~\cite{bib45,bib46}, as both methods learn by comparing statistical differences between distinct equilibrium states or distributions. However, unlike CD, which relies on stochastic sampling~\cite{bib47} to approximate the free distribution, EP constructs gradient estimates using two neighboring equilibrium states of the same system under continuous dynamics~\cite{bib48}, as illustrated in Fig.~\ref{fig:ep_mechanism}. This property endows EP with stronger physical consistency and implementability, making it particularly suitable for training models based on physical energy functions, as well as quantum or quantum-inspired systems.

\subsection{Datasets and Experimental Tasks}
\label{subsec:datasets}

To systematically evaluate the performance and generalization capability of the proposed Quantum Convolutional Neural Network  across visual tasks of varying complexity, we conduct experiments on three standard image classification benchmarks: MNIST~\cite{bib11}, Fashion-MNIST~\cite{bib53}, and CIFAR-10~\cite{bib54}.

The MNIST and Fashion-MNIST datasets consist of single-channel grayscale images with clear local structures and relatively low semantic complexity. These datasets are widely adopted to evaluate model convergence behavior, training stability, and robustness under discrete or binary representations. In contrast, the CIFAR-10 dataset comprises multi-channel natural images with significantly higher spatial complexity and intra-class variability, posing a more challenging benchmark for evaluating representation capacity and generalization performance.

\vspace{0.5em}
Across all datasets, classification is formulated as a supervised learning task with one-hot encoded labels. Dataset-specific preprocessing and normalization follow standard protocols to ensure fair comparison with existing classical and quantum baselines.

\subsection{Baseline Methods and Fair Experimental Design}
\label{subsec:baselines}

To comprehensively assess the effectiveness of the proposed approach, we compare it against a diverse set of representative classical and quantum machine learning models. These baselines span the three dominant paradigms in contemporary quantum learning research: quantum annealing--based methods~\cite{bib15,bib23}, variational quantum circuit--based approaches~\cite{bib52}, and gate-model quantum convolutional neural networks~\cite{bib55}.

\paragraph{(1) Classical Convolutional Neural Network (Classical CNN)}
A classical convolutional neural network is employed as the primary classical baseline. Its network depth, convolution kernel size, number of feature maps, and overall parameter scale are carefully matched to those of the proposed EPIC-CIM. The only differences lie in the mechanism for generating intermediate representations and the training procedure. This design ensures that performance differences can be attributed to the introduction of energy-based quantum optimization and sampling mechanisms, rather than architectural advantages.


\paragraph{(2) Quantum Annealing--Based Methods (QA)}
Quantum annealing approaches, such as those implemented on the D-Wave platform~\cite{bib15,bib23}, formulate learning or inference tasks as the minimization of QUBO or Ising energy functions. These methods are naturally suited for binary variables and combinatorial optimization problems, and do not rely on gate-model quantum circuits. Instead, solutions are obtained through quantum or simulated annealing dynamics. However, due to limitations in hardware connectivity and input encoding, such approaches struggle to directly support convolutional architectures with multiple receptive fields and feature maps.

\paragraph{(3) Variational Quantum Circuit--Based Equilibrium Models (QDEQs)}
We further include Quantum Deep Equilibrium Models (QDEQs)~\cite{bib52} as a representative gate-model quantum baseline. QDEQs leverage variational quantum circuits (VQCs) and replace explicit deep circuit stacking with the computation of stable equilibrium states. While this strategy reduces effective circuit depth, training still relies on parameterized quantum circuits and classical optimization loops, distinguishing it fundamentally from annealing-based or CIM-based approaches.

\paragraph{(4) Gate-Model Quantum Convolutional Neural Networks (QCNN)}
In addition, we compare against gate-model quantum convolutional neural networks ~\cite{bib55}. These methods emulate classical convolution operations through carefully designed quantum circuits with specific symmetries or conservation constraints. They represent a well-established research direction in quantum machine learning, particularly under limited qubit counts and shallow circuit depth.

\subsection{EPIC-CIM Architecture and Training Mechanism}
\label{subsec:cim_qcnn}

Distinct from the above approaches, the proposed EPIC-CIM does not rely on gate-model quantum circuits or variational parameter optimization. Instead, it leverages a Coherent Ising Machine to perform energy minimization and sampling within an energy-based learning framework.

The overall architecture follows a three-stage pipeline: classical convolutional feature extraction, quantum energy-based intermediate representation, and discriminative output generation. Specifically, real-valued convolutional layers are first applied to the input images to extract local spatial features, yielding continuous-valued feature maps. These features are subsequently mapped to external driving terms in the corresponding Ising or QUBO energy function.

The intermediate representation $X_2$ consists of a set of binary variables whose states are not obtained through conventional forward propagation, but rather through sampling from the low-energy distribution of the quantum system. This sampling process captures collective correlations and implicit regularization effects induced by the energy landscape. The output layer $y$ is linearly coupled to $X_2$ for final classification, and is encoded in binary or one-hot form to support supervised learning objectives.

\subsection{Hardware Constraints and Experimental Protocol}
\label{subsec:hardware}

In numerical simulation experiments, model performance is evaluated using the full test set of each dataset. Due to limitations in current quantum hardware---including system scale, connectivity, and operational cost---experiments involving physical quantum systems are conducted using a randomly selected subset of 1000 training samples and 100 test samples.

It is worth noting that, under present hardware conditions, quantum annealing platforms such as D-Wave cannot be readily scaled to full-resolution convolutional neural network training. This limitation arises primarily from hardware connectivity topology, qubit availability, and input encoding constraints, rather than from fundamental algorithmic deficiencies. In contrast, the proposed EPIC-CIM mitigates these issues by decoupling classical convolutional feature extraction from CIM-based energy optimization, enabling scalable and flexible hybrid learning.

\section{Results}\label{sec11}

\subsection{Benchmark Comparison on Standard Datasets}
\begin{table}[]
  \centering
  \caption{Accuracy on the MNIST and Fashion-MNIST Datasets.\footnotesize{\textit{Note:} The QDEQs model does not report training accuracy or standard deviation ($\pm$) in the original paper.}}
  \label{tab:accuracy}
  \begin{tabular}{ccccccc}
    \hline
                              &                & D-Wave     & QDEQs & QCNN       & EPIC-CIM       & CNN   \\ \hline
    \multirow{2}{*}{MNIST}    & Train Accuracy & 93.08±0.25 &   \diagbox{}{}   & 91.33±0.36 &  88.91±0.21             & 97.38 \\
                              & Test Accuracy  & 87.53±0.52 & 73.68 & 84.59±0.91 & \textbf{92.32±0.37} & 97.12 \\ \hline
    \multirow{2}{*}{F-MNIST}  & Train Accuracy & 84.56±0.71 &    \diagbox{}{}   & 82.95±0.47 &   76.23±0.12             & 87.71 \\
                              & Test Accuracy  & 78.96±1.40 & 72.11 & 78.29±0.83 &   \textbf{79.62±0.17}             & 86.47 \\ \hline
    \multirow{2}{*}{CIFAR-10} & Train Accuracy &    \diagbox{}{}      &    \diagbox{}{}   & 35.65±0.43 & 38.21±0.92              & 40.75 \\
                              & Test Accuracy  &  \diagbox{}{}       & 25.45 & 27.79±0.85 & \textbf{34.56±2.01}               & 40.22 \\ \hline
  \end{tabular}
\end{table}

The experimental results on the MNIST, Fashion-MNIST, and CIFAR-10 datasets are summarized in Table~\ref{tab:accuracy}. Overall, the proposed EPIC-CIM demonstrates consistent advantages over existing quantum baseline models across datasets with different levels of visual complexity, highlighting its strong feature learning capability, generalization performance, and optimization stability.

On the MNIST dataset, EPIC-CIM achieves a test accuracy of 92.32\% when trained on the full dataset. Notably, even under limited-data conditions (e.g., using only 2000 training samples), the model still attains an accuracy of 88.64\%, demonstrating strong sample efficiency. It is important to note that all other quantum baseline models are trained on relatively small-scale datasets containing only a few thousand samples. Therefore, achieving superior performance under the same limited-data setting further validates the effectiveness and generalization capability of EPIC-CIM.

Compared with the D-Wave quantum annealing approach (87.53\%), EPIC-CIM improves the accuracy by approximately 5 percentage points. Compared with the standard QCNN (84.59\%), the improvement exceeds 7 percentage points, indicating that the introduced CIM mechanism significantly enhances feature extraction and representation learning. In contrast, QDEQs achieves a relatively lower accuracy of 73.68\%, suggesting potential limitations in optimization or representational capacity under the current experimental setting. These results demonstrate that CIM-inspired optimization can effectively improve the modeling of complex features while enhancing training stability and generalization, particularly in data-scarce scenarios.

The experimental results on the Fashion-MNIST dataset further demonstrate the robustness of EPIC-CIM. Specifically, EPIC-CIM achieves a test accuracy of 79.62\%, outperforming all competing quantum models. Compared with the D-Wave approach, the proposed method improves accuracy by approximately 0.7 percentage points; compared with QDEQs (72.11\%), the improvement reaches about 7 percentage points; and compared with the standard QCNNs (78.29\%), the improvement is approximately 1.3 percentage points. These results indicate that EPIC-CIM maintains strong classification capability even on more challenging datasets with less distinct local features. In addition, the relatively low variance observed during training and testing suggests improved optimization stability compared with other quantum neural network models. This further confirms the robustness and generalization ability of EPIC-CIM across different visual tasks.

To further evaluate scalability on more complex natural image classification tasks, experiments are conducted on the CIFAR-10 dataset. Compared with MNIST and Fashion-MNIST, CIFAR-10 contains richer semantic information, higher intra-class variability, and more complex background distributions, making feature extraction and optimization significantly more challenging for quantum neural network models. Under this setting, EPIC-CIM achieves a test accuracy of 34.56\%, outperforming both QDEQs (25.45\%) and the standard QCNN (27.79\%). Specifically, the proposed method improves the accuracy by approximately 9 percentage points over QDEQs and about 7 percentage points over QCNN, demonstrating its superior representational capability and optimization effectiveness in complex image scenarios.

Although the classical CNN still achieves the highest accuracy (40.22\%), the performance gap between EPIC-CIM and the classical counterpart is substantially reduced compared with other quantum models. This suggests that the introduced CIM mechanism effectively alleviates the optimization difficulties and limited expressivity commonly encountered in quantum convolutional architectures. Furthermore, the relatively stable standard deviation ($\pm2.01$) indicates that EPIC-CIM maintains acceptable training robustness despite the increased complexity of the dataset. These findings demonstrate that the proposed CIM-inspired optimization strategy is not only effective for simple grayscale image datasets, but also scalable to more challenging multi-class natural image classification tasks.

In summary, the experimental results consistently demonstrate the superior performance, robustness, and scalability of EPIC-CIM across different visual tasks. On MNIST, the model achieves high accuracy and strong generalization under both full-data and limited-data settings, highlighting its sample efficiency. On the more challenging Fashion-MNIST dataset, EPIC-CIM continues to deliver stable and leading performance, reflecting its robustness to complex data distributions. On CIFAR-10, the proposed method further demonstrates improved scalability and optimization capability in complex natural image classification scenarios. Overall, these findings provide strong empirical evidence for the effectiveness of integrating coherent Ising machine optimization with quantum convolutional neural networks, and support its applicability to more complex tasks and real-world scenarios.

\subsection{Impact of Training Sample Size}

\begin{figure}[h]
\centering
\includegraphics[width=1\textwidth]{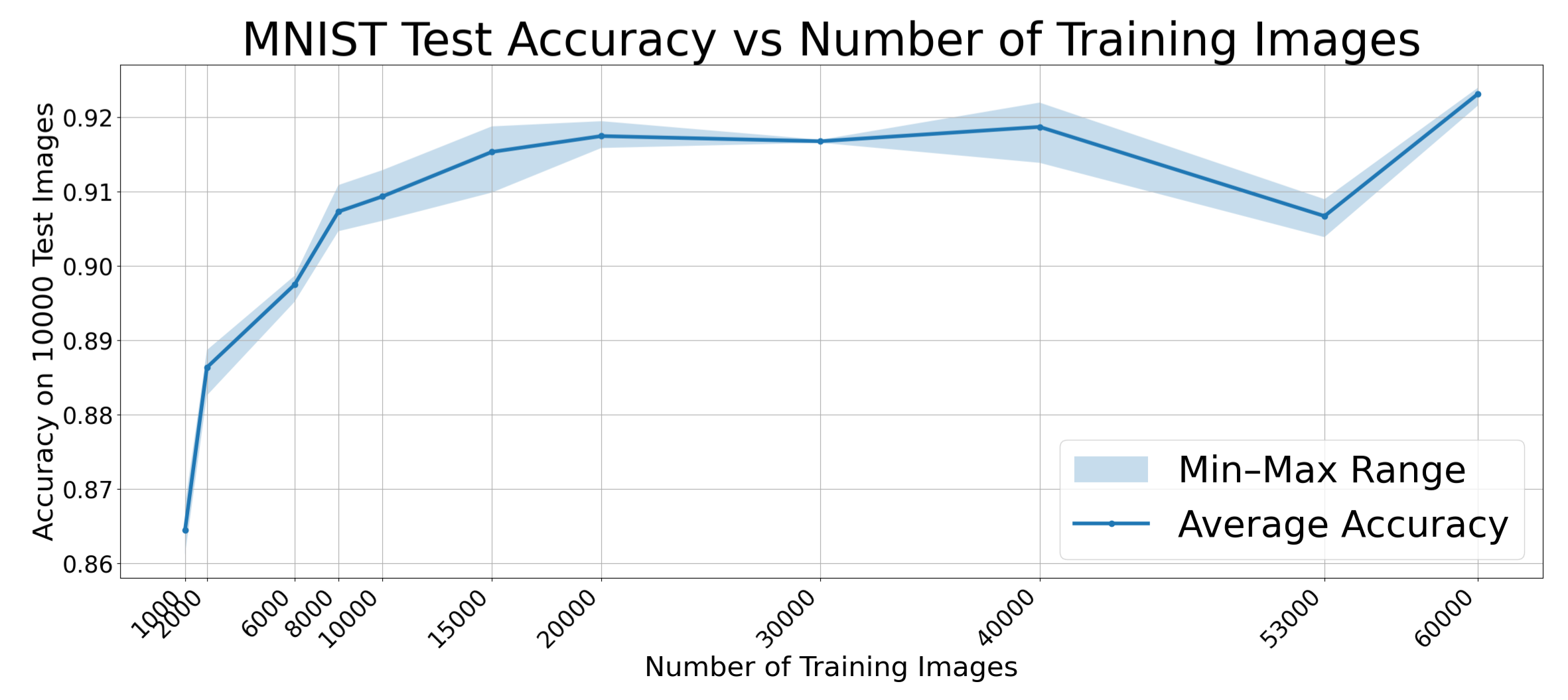}
\caption{Test Accuracy under Different Training Sample Sizes.The results show that the proposed model maintains competitive performance even with limited training samples.}
\label{fig:sample_size}
\end{figure}

Furthermore, this study systematically investigates the impact of training sample size on the performance of the EPIC-CIM model, as illustrated in Fig.~\ref{fig:sample_size}. On the MNIST dataset, as the number of training samples increases from 1,000 to 60,000, the test accuracy exhibits a steady and consistent upward trend, improving from 86.45\% to 92.32\%. Notably, in the low-data regime (1k--8k), the performance gain is particularly significant, indicating that EPIC-CIM is highly sensitive to the amount of training data and can effectively exploit limited samples to achieve rapid performance improvement.

As the training sample size further increases (beyond 20k), the performance gain gradually saturates, suggesting that the generalization capability approaches the capacity limit of the network. This phenomenon highlights that the model can converge quickly under small-sample conditions while maintaining stability in large-scale settings, demonstrating the advantage of the quantum-inspired optimization strategy during training. It is also worth noting that the variance across three independent runs remains low, further confirming the robustness and reproducibility of the training process.

From a broader theoretical perspective, these results are consistent with fundamental principles in statistical learning theory: as the amount of training data increases, the generalization error progressively decreases and the overall performance stabilizes. This behavior not only validates the robustness of EPIC-CIM across different data scales but also demonstrates its strong capability in feature representation and generalization. In other words, the model can efficiently extract key features under data-scarce conditions and maintain stable performance when sufficient data are available, without evident overfitting. These findings provide solid experimental and theoretical support for its application in both low-data and large-scale scenarios, further highlighting the potential of quantum convolutional neural networks in complex visual tasks.

\subsection{Experimental Validation on Quantum Hardware}

To further evaluate the practical feasibility of the proposed EPIC-CIM framework, we conduct experiments on a real quantum hardware platform. Following the sample efficiency analysis in the previous section, we select 1,000 training samples and 100 test samples as a representative low-data setting, which aligns with realistic hardware constraints while ensuring reliable evaluation.


Although the current quantum hardware is subject to resource limitations and operates under a reduced 14-bit precision representation instead of standard 32-bit floating-point precision, the proposed EPIC-CIM still achieves a test accuracy of 91\%. Importantly, this performance remains close to that obtained in high-precision simulations, demonstrating only a minor performance gap.

This result indicates that, despite hardware constraints and limited numerical precision, the proposed model is able to preserve stable representations and decision boundaries. The small discrepancy between hardware and simulation results further confirms the robustness and reliability of EPIC-CIM under realistic implementation conditions, highlighting its potential for practical deployment on near-term quantum and quantum-inspired hardware.

\section{Discussion}\label{sec12}

The experimental results presented in this work consistently demonstrate the effectiveness and robustness of the proposed EPIC-CIM framework across different evaluation settings, including standard benchmark datasets, varying training sample sizes, and real quantum hardware implementations.

From the benchmark comparisons, EPIC-CIM achieves superior performance over representative quantum baseline models, particularly in data-scarce scenarios. The results indicate that the proposed energy-based formulation, combined with quantum sampling and CIM-driven optimization, significantly enhances feature representation and generalization capability, especially compared with variational quantum circuit-based and annealing-based methods.

The analysis of training sample size further reveals that EPIC-CIM exhibits strong sample efficiency. The model is able to rapidly improve performance under low-data regimes and gradually saturates as the dataset size increases, which is consistent with statistical learning theory. This behavior highlights the effectiveness of energy-based learning in stabilizing optimization dynamics and improving convergence properties in discrete parameter spaces.

Importantly, the quantum hardware experiments confirm the practical feasibility of the proposed approach. Despite operating under reduced 14-bit precision and hardware constraints, EPIC-CIM maintains performance close to that of high-precision simulations, with only a minor degradation. This demonstrates that the proposed framework is robust to quantization effects and implementation noise, making it suitable for near-term quantum and quantum-inspired computing platforms.

Overall, these results validate that integrating equilibrium propagation with energy-based quantum convolutional architectures provides a unified and physically interpretable learning paradigm. It enables stable optimization in non-differentiable and discrete systems while preserving strong generalization performance. Future work will focus on scaling the framework to deeper architectures, improving hardware-efficient implementations, and exploring broader applications in large-scale vision and combinatorial optimization tasks.

\section{Conclusion}\label{sec13}

This work systematically evaluates the performance and generalization capability of the proposed EPIC-CIM under varying training sample sizes, visual task complexities, and real quantum hardware conditions. The experimental results demonstrate that the model achieves rapid performance improvement in low-data regimes, exhibiting strong sample efficiency and the ability to effectively learn discriminative features from limited data. As the training data scale increases, the performance gradually stabilizes, indicating robust generalization and stable training dynamics in large-scale settings. This behavior is consistent with fundamental principles in statistical learning theory, further validating the robustness of the proposed model across different data regimes.

The EPIC-CIM framework enhances feature extraction and representation learning by incorporating a quantum-inspired optimization module. During training, the model captures collective correlations among input features and introduces an implicit regularization effect through energy-driven low-energy state sampling. This mechanism enables fast convergence under limited data conditions while maintaining stable performance in large-scale scenarios. Compared with conventional approaches based on variational quantum circuits or gated quantum convolution, the proposed method achieves more efficient and flexible training, even under constrained computational and hardware resources.

Importantly, the effectiveness of the proposed framework is further validated on real quantum hardware. Despite operating under reduced precision constraints (14-bit representation) and inherent hardware noise, the EPIC-CIM maintains performance close to that of high-precision simulations, with only a minor performance gap. This result demonstrates that the proposed energy-based learning formulation is robust to quantization effects and physical implementation limitations, confirming its feasibility for deployment on near-term quantum and quantum-inspired devices.

Overall, the experimental results confirm the superior performance and robustness of EPIC-CIM across different data scales, task complexities, and hardware environments. The model not only demonstrates strong learning capability in data-scarce scenarios but also maintains stable and reliable performance in high-dimensional, multi-class, and complex visual tasks, as well as in real quantum hardware implementations. These findings highlight the effectiveness of integrating quantum-inspired optimization with convolutional neural networks, and provide strong empirical and theoretical support for its application in resource-constrained quantum hardware environments. Furthermore, this work underscores the potential of quantum convolutional neural networks as a scalable and practical framework for future intelligent systems.

\backmatter

\bibliography{sn-bibliography}

\end{document}